# Super-resolved optical mapping of reactive sulfur-vacancy in 2D transition metal dichalcogenides

Miao Zhang[1,2,*], Martina Lihter[1], Michal Macha[1], Karla Banjac[3], Yanfei Zhao[4,5], Zhenyu Wang[4,5], Jing Zhang[4,5], Jean Comtet[1], Magalí Lingenfelder[3], Andras Kis[4,5] and Aleksandra Radenovic[1,*]

[1]Laboratory of Nanoscale Biology, Institute of Bioengineering, School of Engineering, École Polytechnique Fédérale de Lausanne (EPFL), 1015 Lausanne, Switzerland.
[2]Department of Applied Physics, KTH Royal Institute of Technology, Stockholm, Sweden
[3]Max Planck-EPFL Laboratory for Molecular Nanoscience, and Institut de Physique, École Polytechnique Fédérale de Lausanne (EPFL), Lausanne, Switzerland.
[4]Electrical Engineering Institute, École Polytechnique Fédérale de Lausanne (EPFL), Lausanne, Switzerland.
[5]Institute of Materials Science and Engineering, École Polytechnique Fédérale de Lausanne (EPFL), Lausanne, Switzerland.

[*]miao.zhang@epfl.ch, aleksandra.radenovic@epfl.ch

**Transition metal dichalcogenides (TMDs) represent an entire new class of semiconducting 2D materials with exciting properties. Defects in 2D TMDs can crucially affect their physical and chemical properties. However, characterization of the presence and spatial distribution of defects is limited either in throughput or in resolution. Here, we demonstrate large area mapping of reactive sulfur-deficient defects in 2D-TMDs coupling single-molecule localization microscopy with fluorescence labeling using thiol chemistry. Our method, reminiscent of PAINT strategies, relies on the specific binding by reversible physisorption of fluorescent probes to sulfur-vacancies via a thiol group and their intermittent emission to apply localization of the labeled defects with a precision down to 15 nm. Tuning the distance between the fluorophore and the docking thiol site allows us to control Föster Resonance Energy Transfer (FRET) process and reveal large structural defects such as grain boundaries and line defects, due to the local irregular lattice structure. Our methodology provides a simple and fast alternative for large-scale mapping of non-radiative defects in 2D materials and paves the way for in-situ and spatially resolved monitoring of the interaction between chemical agent and the defects in 2D materials that has general implications for defect engineering in aqueous condition.**

The emerging class of two-dimensional (2D) materials, such as graphene, hexagonal boron nitride (h-BN) and transition metal dichalcogenide (TMD) monolayers show exciting physical properties that are distinct from their bulk forms. However, experimentally measured values of these properties, such as electron mobility, photoluminescence (PL), dielectric screening are largely affected by the presence of defects that are introduced unintentionally in the monolayers during material growth and processing. Although for the many practical applications of semiconductor devices based on 2D materials defects are thus often not desired, novel properties induced by defects can also be exploited. To name a few, hydrogen evolution reaction can be improved by defects engineered $MoS_2$[1,2]; single photon emitters have been discovered in $WSe_2$ and hBN on defects-related structures[3,4]; magnetism has been introduced by metal vacancies in $PtSe_2$[5]; gate-tunable memrisitive devices have been built based on grain boundaries in $MoS_2$[6,7]. Due to the atomic-thickness of the 2D materials, defects in such hosting material are highly sensitive to their environment, and can be inversely used as sensing probes[8]. It is therefore of utmost interest to study and control defects in 2D materials. Especially, with the rapid progress in the large scale 2D material growth by metalorganic chemical vapor deposition (MOCVD)[9], there is a growing demand for a fast, large-scale characterization method of the presence and distribution of individual defects in 2D materials.

Mapping of defects in 2D materials is often a delicate trade-off between the imaging resolution and the imaging scale. Atomic resolution is often achieved within tens of nanometers imaging area by various electron microcopies[10–13], whereas optical microscopy and spectroscopy can reveal diffraction-limited defect spatial distribution in relatively large area (tens of micrometers)[3,14,15]. To bridge this gap, we recently demonstrated the application of single molecule localization microscopy (SMLM) for wide-field mapping of room temperature emitting defects in hBN with resolution down to 10 nm[16,17]. However, for the most abundant point defects in TMDs, e.g. the sulfur vacancies in $MoS_2$ and $WS_2$, SMLM cannot be directly applied due to their non-radiative nature at room temperature[10,15,18].

An alternative route to achieve super-resolution microscopy is point accumulation for imaging in nanoscale topography (PAINT) and its variations[19–21]. Here, free diffusing fluorescent probes transiently bind to the target site resulting in sparse stochastic blinking that allow single molecule localization with high precision (<10 nm)[22]. With chemical engineering of the probe, interaction towards specific targets can be achieved.

Coupling PAINT methods with thiol chemistry, here we demonstrate visualization of the spatial distribution of non-radiative defects in 2D materials, using thiol-functionalized fluorescent probes that specifically interact with sulfur-vacancies[23–27]. The methodology is abbreviated as 2D-PAINT here after. By direct observation, we establish the binding affinity and specificity of the reversibly physisorbing thiol-functionalized probes and reveal correlation between the defect density and the PL emission of the TMD monolayers. By varying the distance between the fluorophore and the docking thiol group, we can further control Föster resonance energy transfer (FRET) induced fluorescent quenching[28,29] and reveal fine details of large structural defects, such as grain boundaries and line defects. Our work open perspectives for mapping of a broad range of non-radiative defects in 2D materials and pave the way for in-situ and spatially resolved monitoring of the interaction between chemical agent and defects in 2D materials, that can potentially lead to defect healing, 2D material modification, as well as bio-sensing applications.

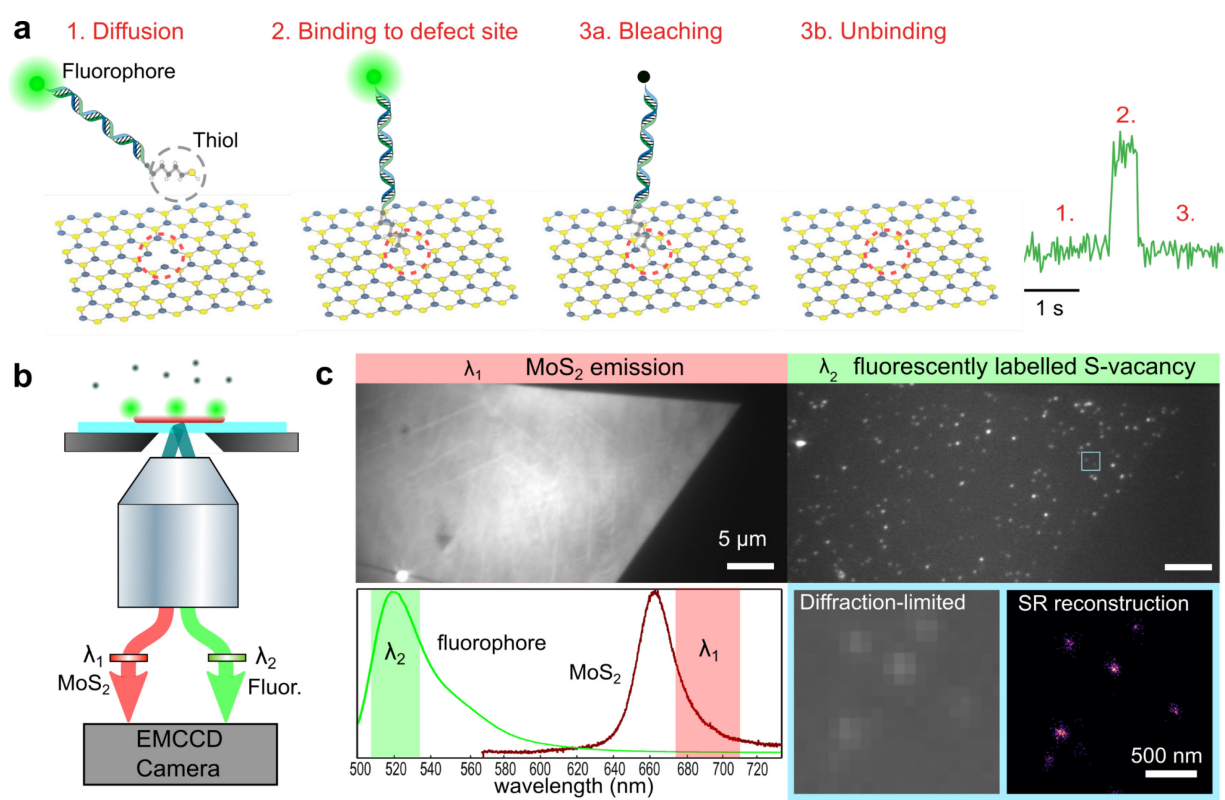

**Figure 1. Working principle of thiol-chemistry-assisted super-resolution optical mapping of sulfur-vacancies in 2D transition metal dichalcogenides.** **(a)** Fluorophores with emission wavelength clearly distinguishable from the photoluminescence (PL) of $MoS_2$ are conjugated with dsDNA linker molecules and docking thiol molecules (1). The selective adsorption of thiol molecules at sulfur-vacancies allows labeling of these defects. (2). The dynamic of the adsorption/desorption of thiol molecules (3a) or the bleaching of fluorophores (3b) yield intermittent fluorescent signal from the fluorophores bound to a defect site, which allows the application of single-molecule localization microscopy. **(b)** The optical setup is based on total internal

reflection fluorescent microscopy. The evanescent wave of the laser beam confines excitation and photoluminescence (PL) emissions from MoS$_2$ and fluorophores to ~100 nm from the surface, collected by a 100 × NA 1.49 oil immersion objective lens. The emission signals from MoS$_2$ and the fluorophores are then split into two paths according to their wavelength and projected side-by-side on an EMCCD camera (see Figure S1 for setup) **(c)** PL images of a MoS$_2$ flake (path λ$_1$) and fluorescently labelled sulfur vacancies (path λ$_2$) are collected simultaneously. The images were taken with ATTO-dsDNA70bp-SH probes with a concentration of 100 pM under excitation power of ~15 W/cm$^2$. Images shown here are averaged over 1000 frames for clarity. The bottom panel shows the spectra of ATTO-488 and transferred MoS$_2$. The spectrum ranges of the bandpass filters are marked in green and red. PL image (standard deviation of 10$^4$ frames) of the zoomed-in area highlighted in blue-square showing diffraction-limited fluorescent pattern due to individual thiol molecules. After localization of the position of each molecule, overlay of the centroid positions of the detected molecules in each frame yields the reconstructed super-resolution (SR) map.

**Principles of 2D-PAINT and experimental conditions**

The principle of 2D-PAINT is depicted in Figure 1. To enable detectable transient labeling of sulfur-vacancies, we design probes that each consists of a fluorophore, an alkane thiol molecule and a double-stranded (ds) DNA as a spacer inbetween to control FRET induced fluorescent quenching (Figure 1a). A total internal reflection fluorescence (TIRF) microscope is used to image labelled sulfur-vacancies in MOCVD-grown monolayer of MoS$_2$ flakes with a minimized PL background from the free diffusing probes in solution. Prior to imaging, MoS$_2$ flakes are transferred to glass cover slips [30]. The fluorophores ATTO488 and fluorescein amidite (FAM) with emission peak at 520 nm are selected to avoid overlapping with PL emission of MoS$_2$, which is typically between 590 nm – 720 nm (i.e. 1.7 eV and 2.2 eV)[31,32] (Figure 1c). A laser with wavelength of 488 nm excites both the fluorophores and the MoS$_2$ through an oil immersion TIRF objective lens. The PL image is split into two paths corresponding to the fluorophore and MoS$_2$ emission respectively, as shown in Figure 1b, c. This simultaneous imaging allows us to correlate the defect distribution and the photoluminescence of the 2D-TMDs. A high concentration salt buffer (400 mM KCl 40 mM Tris buffer pH~8) is used to shield the electrostatic repulsion between the negatively charged MoS$_2$ surface and the negatively charged DNA molecules. In the absence of the oxygen-scavenger buffer, fluorescent probes that attach to the surface bleach rather quickly under the continuous excitation of the 488 nm laser, such that only a fraction of the attached probes is detectable on each frame allowing for localization. As can be seen in the zoomed-in images in Figure 1c, PL signal from a single fluorescent thiol label spreads over ~5-by-5 pixels (pixel size = 105 nm) according to the Point Spread Function (PSF) of the optical system. The centroids of the molecules imaged in each frame are localized by 2D-Gaussian fitting with the localization precision that scales inversely with square root of total number of emitted photons[33]. We obtain the best precision of ~15 nm with ATTO488 fluorophore under excitation power of 200 W/cm$^2$. Detailed description of the localization procedure can be found in Methods.

**Binding affinity and specificity**

First, we perform 2D-PAINT imaging on MoS$_2$ flakes using fluorescent probes consist of FAM dye, dsDNA of 70 base pairs and a thiol functional group, namely FAM-dsDNA70bp-SH. The schematics of the probe, with the resulting PAINT image is depicted in Figure 2a. The localizations based on the detected fluorescence of individual emitters are summed up from $10^4$ frames. Brighter spots represent a higher localization density. As shown in Figure 2a, with FAM-dsDNA70bp-SH probes, bindings are homogeneous over the sample with a slightly higher density of events along the edges, as shown in Figure 2a. This binding homogeneity is consistent with the homogeneous defect distribution expected from TEM and PL measurements on perfect triangle shape flake with homogeneous PL (see in Figure 2b). Straight edges were shown previously to be Mo-terminated[34], thereby are more likely to be bound by thiol probes.

As a control experiment, we employ probes without thiol, namely FAM-dsDNA70bp. All other conditions are kept identical. As shown in Figure 2c, in contrast to the high localization density with thiol probes, only a few fluorescent events are detected when the probe is switched to FAM-dsDNA70bp. Such a marked difference proves that -SH is the active group responsible for the interaction between the probes and the MoS$_2$ surface that yields the PL signal. AFM scanning of as-grown MoS$_2$ on sapphire in buffer solution with the presence of thiol probes shows that the molecules can also physisorb with their main molecular axis parallel to the surface (Figure S11). However, such adsorption geometries not interacting directly at the defects will not contribute to the fluoresence signal due to the quenching by MoS$_2$ at close proximity[28,29]. Indeed, using fluorophores that are directly conjugated to ethanethiol molecules (ATTO-SH), we observe almost no fluorescence on the MoS$_2$ flake, as shown in Figure 2d. These results demonstrate the necessity of the thiol functional group and the dsDNA linker molecule to enable fluorescence detection of the probe-MoS$_2$ interaction.

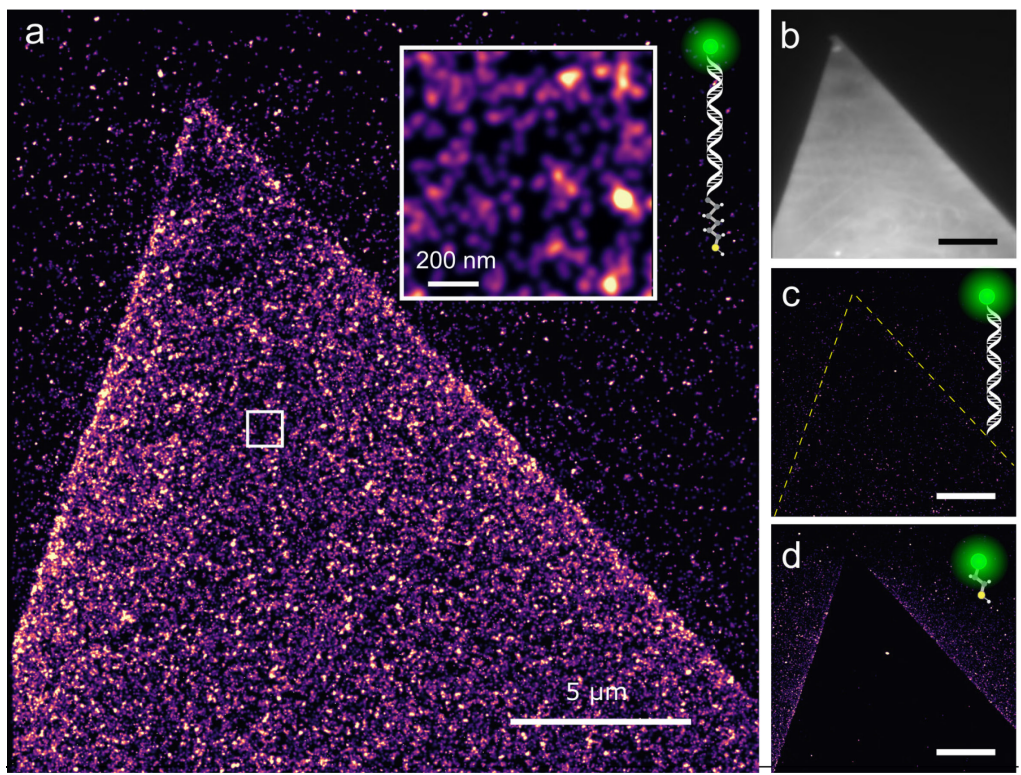

**Figure 2. Affinity of thiol binding on MoS$_2$ surface: 2D-PAINT reconstructed images of a MoS$_2$ flake with fluorescent probes of different compositions.** **(a)** 2D-PAINT image acquired using fluorescent probes consisting of a fluorophore head, dsDNA of 70 bp as linker molecule, and a thiol tail (FAM-dsDNA70bp-SH). The zoomed-in image reveals a high density of binding on MoS$_2$. **(b)** An averaged PL image of the MoS$_2$ flake in (a) with a perfect 60° triangle shape and homogeneous PL intensity. **(c)** 2D-PAINT image acquired using fluorescent probes without thiol groups (FAM-dsDNA70bp). **(d)** 2D-PAINT image acquired using fluorophores that directly conjugated with thiol (ATTO-SH). All experiments are performed with 40 nM concentration of the probes. 2D-PAINT images are reconstructed from $10^4$ frames with normalized Gaussian rendering. Scale bars in b-c: 5 µm.

To validate the binding specificity on defects, we perform 2D-PAINT imaging on MoS$_2$ samples with pre-patterned defective sites. High density of defects were deliberately introduced in MoS$_2$ flakes by Xenon focused-ion beam (FIB) irradiation[35]. Periodical patterns with a pitch distance of 2 µm are irradiated on MoS$_2$ with carefully tuned ion dose and dwell time to avoid complete removal of the material, which is confirmed by the Raman scattering spectra, as shown in Figure 3c and AFM (Figure S2). On irradiated samples, we observe a downshift of E' mode and an upshift of A'$_1$ mode with a clearly decreasing of the amplitude ratio of E'/A'$_1$ and broadening of both peaks. In addition, disorder-induced modes emerge as the dose of irradiation increases, one of which is at the low-frequency shoulder of E' (~377 cm$^{-1}$) and the other LA(M) mode at ~227 cm$^{-1}$ (Figure S3)[36]. These transitions in Raman scattering spectra strongly indicate an increasing concentration of sulfur-vacancies, which has been reported previously upon ion or electron irradiation[14,36,37]. The highly defective crystal structures induced by the ion bombardment lead to a diminished PL emission of $MoS_2$, as shown in Figure 3a (gray scale). Similar PL quenching has been reported on $MoS_2$ with distorted lattice after oxygen plasma exposure[37]. Performing 2D-PAINT imaging on the flake using ATTO-dsDNA70bp-SH probes, we observe a highly selective binding on the irradiated area. Overlaying the reconstructed localization density map (color-coded) with the PL image of the same area demonstrates a strong correlation between the area of diminished PL and the area of high localization density. Details of the overlay can be seen in the zoomed-in image in Figure 3b. The absence of binding in the non-irradiated area is most likely due to re-deposition of the knocked-off atoms that blocks the binding sites. The correlation between the localization density and the defect density is also confirmed by performing 2D-PAINT imaging on monolayer MOCVD-grown $WS_2$ flakes. MOCVD-grown $WS_2$ is known for its inhomogeneity in PL emission due to the defect density variation across a flake[15,38,39]. 2D-PAINT imaging reveals a strong correlation between the localization density of the fluorescent events and the PL intensity of the $WS_2$ flake (Figure S4). These results clearly demonstrate favorable binding between a thiol probe and a sulfur-vacancy.

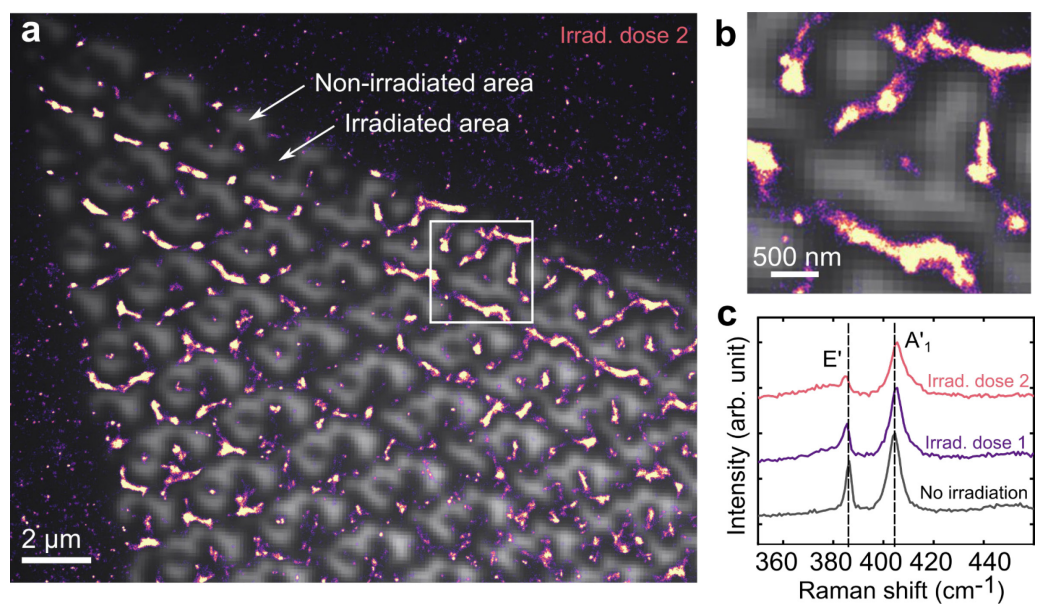

**Figure 3. Specific binding of thiol probe on sulfur vacancy defects: 2D-PAINT imaging on FIB pre-patterned $MoS_2$.** **(a)** Overlay of a PL image (gray scale) and a 2D-PAINT reconstructed image (color-coded) taken on a pre-patterned $MoS_2$ flake by the focused-ion beam at dose of $1.1 \times 10^{14}$ ions/cm$^2$. Scale bar is 2 µm. 2D-PAINT image is reconstructed from $5 \times 10^3$ frames with averaged shifted histogram rendering. Imaging was performed with ATTO-dsDNA70bp-SH probes with a concentration of 10 nM under excitation power of ~56 W/cm$^2$. **(b)** Zoomed-in image of the highlighted area in (a). Scale bar is 500 nm. **(c)** Raman scattering spectra of $MoS_2$ flakes with no irradiation, irradiation dose 1 of $3.4 \times 10^{13}$ ions/cm$^2$, and dose 2 of $1.1 \times 10^{14}$ ions/cm$^2$.

**Non-homogeneous FRET reveals larger structural defects**

The quenching effect evidenced in Figure 2d shows the necessity of the linker molecule. To study the influence of the linker molecule length on the detection of the crystal defects, we designed fluorescent thiol probes of various sizes. Taking advantage of the high spatial resolution of 2D-PAINT, we perform the imaging on MoS$_2$ monolayer crystals that contain large structural defects, such as grain boundaries and line defects to investigate the homogeneity of the FRET effect. Sulfur-deficient grain boundaries and line defects often appear dark in the PL image as shown in Figure 4a[34]. Figure 4b-d show the 2D-PAINT images reconstructed from data taken with probes of 70 bp, 50 bp and 30 bp dsDNA, respectively, corresponding to the contour length of 24 nm, 17 nm and 10 nm. The grain boundaries and line defects become clearly visible with the decreasing probe length. This even applies to the thiol probes without dsDNA linker molecules (Figure S5). Fine details of these large structural defects are revealed beyond the diffraction-limitation in the PL image. We attribute this non-homogenous localization density to distinct FRET quenching processes in areas with and without large structural defects.

We extracted the average binding density per frame on areas with and without large structural defects. As can be seen in Figure 4e, the binding densities on the large structural defects are higher than on the areas without the large structural defects and follow opposite trends with respect to probe length. These opposite trends in binding density are responsible for the improving contrast of the grain boundaries and the line defects in Figure 4d.

The increasing binding density on the large structural defects is consistent with an increase in the diffusion coefficient of the probe in the absence of any FRET quenching effect. In the diffusion-limited regime, the number of arriving probes on each frame is proportional to its diffusion coefficient. Shorter probes diffuse faster than the longer ones[40], therefore having a higher binding density. Similar trend is observed on glass substrate (not shown), which is consistent with the negligible FRET processes in these areas.

That this trend is opposite on areas without large structural defects can be understood based on non-radiative FRET process that reduces the PL intensity of probes, resulting in a cut-off for the low-intensity events in detection. Such truncation is more pronounced for shorter probes. As can be seen in Figure 4f, the fluorophore intensity measured on MoS$_2$ flakes without large structural defects increases with increasing length of dsDNA linker molecule. The theoretical calculation shows that the effective range of FRET effect is about 40 nm (Inset in Figure 4f), matching with the range of dsDNA lengths used in our experiments.

As shown in Figure 4g, the intensity of detected binding events varies indeed according to the crystal structure of MoS$_2$ monolayer. Grain boundaries and line defects in MoS$_2$ have been reported as arrays of dislocations formed by atomic-ring structures at the intersection of two single-crystalline structures with different orientations[11,34,41]. The significantly reduced FRET rate on such defects could originate from disruption of the periodicity of the crystalline structure that changes the local dipoles of the MoS$_2$ crystal, thus hindering the electromagnetic wave coupling between the fluorophore and the MoS$_2$. In addition, strain around the boundaries can modify local band structures, thus affecting the absorption efficiency of certain energy[42,43]. However, further investigation is required to validate this hypothesis which is beyond the scope of this work.

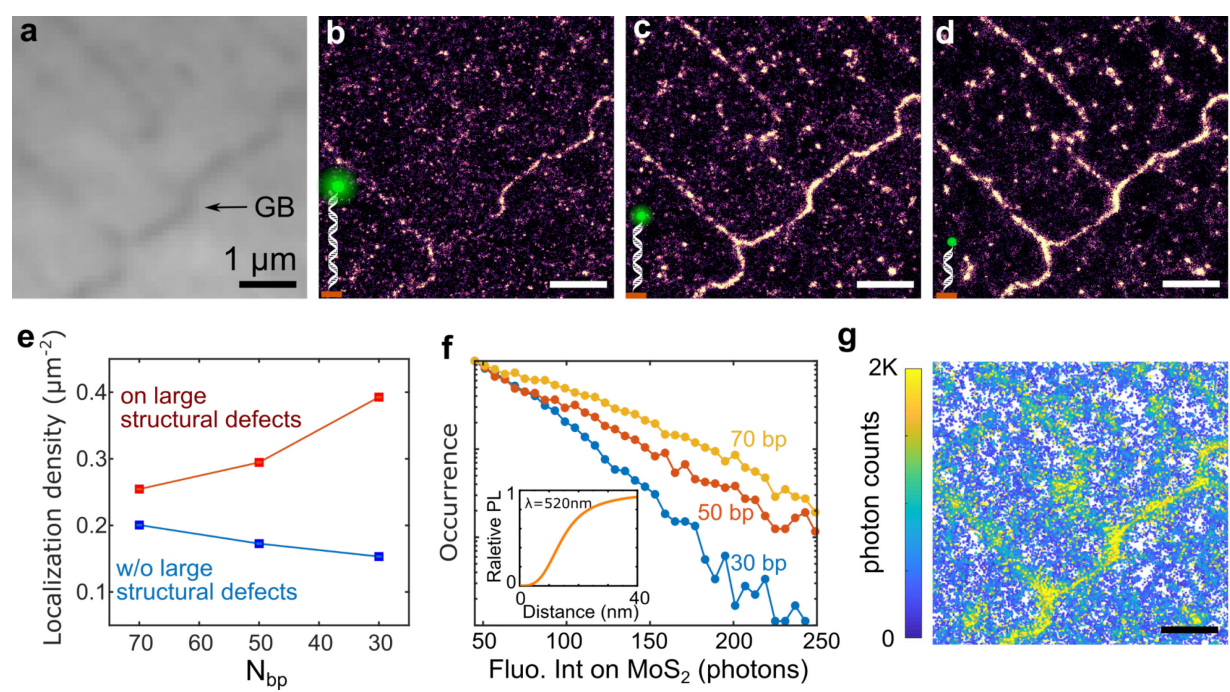

**Figure 4. Fluorescent probes with varied DNA linker lengths reveal large structural defects on MoS$_2$ (a)** An image of PL emission of monolayer-MoS$_2$ consisting of large structural defects, including a grain boundary (GB) and line defects. **(b) - (d)** 2D-PAINT images taken on the same area of MoS$_2$ flake using fluorescent thiol probes ATTO-dsDNA-SH with dsDNA length of 70 bp, 50 bp and 30 bp, respectively, at 40 nM. The excitation power density is 67 W/cm$^2$. The images are reconstructed from 5×10$^3$ frames using averaged shifted histogram. **(e)** Detected localization density per frame as a function of the number of base pairs in DNA linkers. Data are extracted respectively on areas with (red) and without (blue) large structural defects. **(f)** Intensity histogram of ATTO-dsDNA-SH with dsDNA length of 70 bp, 50 bp and 30 bp, respectively, on a MoS$_2$ flake without large structural defects. The excitation power is 19 W/cm$^2$. The inset shows theoretically calculated normalized intensity of the fluorophore ATTO488 emission as a function of the fluorophore distance from the MoS$_2$ surface. Details of the calculation can be found in Supplementary Information. **(g)** Fluorescence intensity map of (d). Scale bars 1μm.

**Binding kinetics**

An important remaining question is the nature of the physical interaction responsible for the binding of the thiol probe on surface defects, for both designing more efficient 2D-PAINT imaging as well as for the future development of defect healing agents. We thus investigated the binding kinetics of thiol probes on MoS$_2$ with bulk concentrations ranging from 100 pM to 100 nM. The average localization density per frame is extracted over $10^4$ frames from the same region of interest of 10-by-10 µm$^2$ selected in the middle of the illuminated MoS$_2$ flakes. As shown in Figure 5a, the localization density increasing with the concentration of thiol probes resembles Langmuir absorption isotherm, which indicates a reversible binding process.

In steady-state, the total number of defects occupied by fluorescent probes reaches a constant value, with a balance between the association and the dissociation of the free diffusing fluorescent probes to the defects and the bleaching of the fluorescent probes bound to the defects. A schematic of the binding process is shown in the inset of Figure 5a. In the case of a reversible binding process, the density of detected fluorescent probes bound to the defects at the equilibrium state is a function of the probe concentration in solution, which can be described mathematically as

$$q = \eta Q \frac{C}{C + \frac{k_d}{k_a}} \frac{k_d}{k_d + k_b} \quad , \quad (1)$$

where $\eta$ is the detection efficiency, $Q$ is the density of total defect sites available for binding, $C$ [M] the concentration of thiol probes in the solution, and $k_d$ [s$^{-1}$] denotes the dissociation rate, $k_a$ [M$^{-1}$·s$^{-1}$] the association rate, $k_b$ [s$^{-1}$] the bleaching rate of the fluorophore. Note that equation (1) is just a classic Langmuir binding equation with an additional factor $\frac{k_d}{k_d + k_b}$, which describes the fraction of unbleached defect-bound probes. The dissociation constant defined as $K_D = \frac{k_d}{k_a}$ [M] is used to characterize the affinity between the probe and the defect. Detailed derivation can be found in the SI. Sec5. Fitting equation (1) to the experimental data, we extract $K_D$ as 6.48 ± 0.48 × 10$^{-9}$ M at a salt concentration of 400 mM with pH ~8. The corresponding molar Gibbs free energy change $\Delta G^o = RT \log(K_D/C^0)$ is ~ -47 kJ·mol$^{-1}$ (19.3 k$_B$$T$) at room temperature (calculation in SI.Sec6). The value is not significantly affected by the choose of fluorophore and the excitation power density up to 40 W/cm$^2$ (SI. Sec 8). This low value of $\Delta G^o$, along with the reversible Langmuir adsorption isotherm, implies that the interaction between a thiol and an S-vacancy is most likely a physical adsorption mediated by an electrostatic interaction.

By analyzing the blinking statistics of each localization spot, we can directly extract the association rate $k_a$ from the reciprocal of the mean OFF time (or the interevent time) via $1/\tau_{OFF} = k_a C$ in the diffusion-limited regime and the sum of the bleaching rate and the dissociation rate from the reciprocal of the mean ON time (event duration time) via $1/\tau_{ON} = k_b + k_d$. As shown in Figure 5b, $1/\tau_{OFF}$ scales indeed linearly with the probe concentration, giving the association rate $k_a = 2.86 \times 10^6$ M$^{-1}$s$^{-1}$ for FAM-dsDNA70bp-SH at a salt concentration of 400 mM KCl. On the other hand, the ON time does not depend on the probe concentration significantly, giving $1/\tau_{ON} = k_b + k_d \sim 3$ s$^{-1}$ with the excitation power density of 5 W·cm$^{-2}$. From $k_d = K_D k_a$, we can extract $k_d \sim 2 \times 10^{-2}$ s$^{-1}$. This implies that a thiol probe stays at a defect site for ~ 50 s on average. The detectable ON time in our experiments is mainly determined by the bleaching of the fluorophore, which is consistent with the photophysics of the fluorophore embedded in 1% polyvinyl alcohol (see in Figure S7). The fraction of the defect bound FAM-dsDNA70bp-SH probes remain fluorescent ($\frac{k_d}{k_d + k_b}$) is only about 0.7%. Given that the detectable binding saturation density in Figure 5a as 0.04 μm$^{-2}$, we can estimate the density of the active sulfur vacancies that can interact with thiol probes in water simutanously is in the order of 10 μm$^{-2}$. This is orders of magnitude lower than the defect density measured with TEM (10$^5$ μm$^{-2}$), although the TEM obtained defect density is reported be higher than the actual value in the material since defects are often as well created by electron beam during TEM imaging[44]. We speculate that this discrepany comes from the deactivation of sulfur vacancies by multiple sources including ~ 20% polymer contamination coverage of MoS$_2$ from the transfer process (Figure S10), surface trapped airbubbles (Figure S11), different charge states of sulfur-vacancy[45], proton and O$_2$ passivation[46,47]. The latter three are reversible processes. Using balanced super-resoltion optical fluctuation imaging (bSOFI) to analyze the chemically active defect density as a function of imaging frames[48], we observe that the density evaluation reaches stable estimation at 16 000 frames (Figure S12). The implies that by this time all chemically active defects have been visited by the probes. We therefore extract the total chemically active defect density at the stable stage as ~2000 μm$^{-2}$.

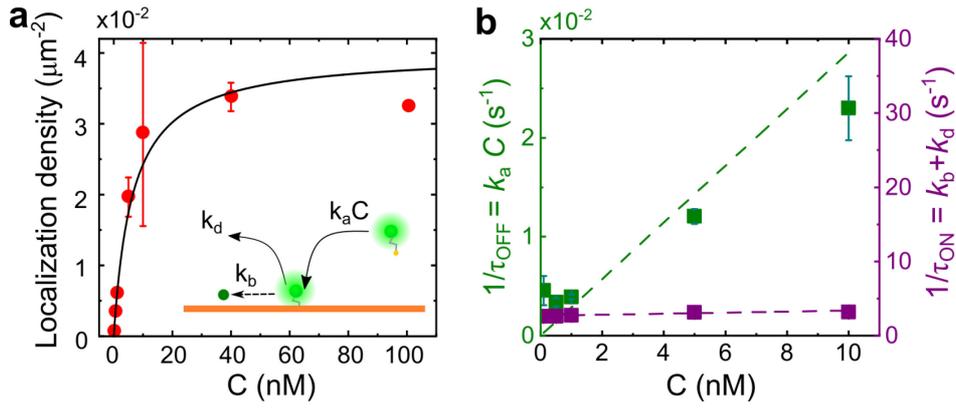

**Figure 5. The Langmuir adsorption isotherm and blinking statistics revealing the binding kinetics between thiol and sulfur vacancies. (a)** The Langmuir adsorption isotherm of FAM-dsDNA70bp-SH binding to a $MoS_2$ monolayer indicates a reversible binding process. The inset shows a schematic of the reversible binding process with the bleaching process of the fluorophore. $k_a$ denotes for the association rate, $k_d$ for the dissociation rate, and $k_b$ for the bleaching rate of the fluorophore. The data points (red dots) is fitted by equation (1). Error bars are standard deviations. **(b)** The reciprocal of the blinking interval time ($1/\tau_{OFF} = k_a C$) and the blinking on time ($1/\tau_{ON} = k_b + k_d$) as a function of the probe concentration (C). The linear fit gives the association rate $k_a$ of $2.86 \times 10^6$ $M^{-1}s^{-1}$ at a salt concentration of 400 mM KCl. The blinking ON time is independent of the probe concentration.

**Discussion and conclusion**

Combining PAINT methodology with defect-specific fluorescent probes allow us to visualize non-radiative defects in 2D materials with resolution beyond the diffraction limit and with a high-throughput at room temperature. This wide-field rapid mapping technique gives a global picture of the distribution of sulfur-deficient point-defects as well as large structure defects across whole flakes of $MoS_2$ and $WS_2$. 2D-PAINT fills the gap between the electron scanning microscopy and diffraction-limited far-field optical methods, with clear advantages of simplicity, fast-scan and throughput compared to device-based defect characterization. By giving unprecedented insight on the interaction between functional groups and target defects, through analysis of single-molecule blinking statistics, it could serve as a platform to study surface chemistry on 2D materials. Moreover, our demonstration of the distance-dependent FRET with the defect-anchored probes on $MoS_2$ on single-molecule level opens up a new perspective in developing 2D-PAINT towards a three-dimensional super-resolution methodology using a similar principle of metal-induced energy transfer, which has been recently demonstrated on graphene[49]. Moreover, our method could be applied to other types of non-radiative defects by versatile chemical modification of the fluorescent probes, such as $WSe_2$. By bridging the gap between super-resolution microscopy and 2D materials, our work is thus an important step towards the multidimensional characterization tool that will facilitate defects engineering in 2D materials and propel the development of a variety of bio-sensing and chemical modification studies with 2D materials.

**Material and methods**

MOCVD growth

Monolayer $MoS_2$ is grown on C-plane sapphire substrates in a home-built system using metal organic chemical vapor deposition (MOCVD) method. As previously described, sapphire substrate is annealed at 1000 °C (for 2h in air) to achieve an atomically smooth surface for epitaxial growth[50]. Before growth, NaCl solution is spin coated on the substrate to suppress nucleation and promote the growth[9,51]. The two precursors $Mo(CO)_6$ and $H_2S$, with the flow rate ratio of 1:6028, are carried by Ar gas to the MOCVD chamber and undergo reaction at 820 °C for 30 min. $Mo(CO)_6$ is kept at 15 °C in a water bath and the valve is closed immediately after growth process, while $H_2S$ continues flowing during the cooling process. Throughout the whole growth process, the furnace is kept at 850 mbar pressure.

Fluorescent thiol probes

Single stranded (ss) DNA with 5' modification of thiol C6 and 3' modification of ATTO-488 or FAM dyes were annealed in-house with complementary strands to form dsDNA probes: ATTO-dsDNA-SH and FAM-dsDNA-SH. ssDNA with 3' modification of FAM were annealed with complementary strands to form dsDNA probes: FAM-dsDNA. Three lengths of DNA were used, namely 30bp, 50bp, and 70bp. All ssDNA products are ordered from MicroSynth. ATTO-SH were ordered from ATTO-TEC.

Microscope setup

Imaging was carried out on a custom-built microscope that was described previously[16]. A schematics of the optical setup is depicted in Figure S1. Briefly, a coverslip with transferred 2D-TMDs is placed on an inverted microscope (IX71, Olympus) with a piezo stage (Nano-Drive, Mad City Labs) driven with a feedback loop to minimize the drift in z-direction. A 100 mW 488 nm laser (Sapphire, Coherent) is used to excite the sample at an angle beyond the critical angle of glass/water boundary (58.9°). The excitation power is controlled by an acousto-optic tunable filter (AOTFnC-VIS-TN,AA Opto-Electronic). The laser beam is focused at the back focal plane of the objective (UApo N ×100, NA 1.49, Olympus) to enable a wide-field illumination. The emission from the sample is collected by the same objective lens through a dichroic mirror (ZT488/561rpc, Chroma) and an emission filter (StopLine 405/488/568, Semrock). The emission goes further through a pre-calibrated adaptive optics (micAO 3DSR, Imagine Optics) to minimize distortion of the point spread function. The emission of $MoS_2$ and fluorophores are split into two paths with wavelength windows of 509 nm-530 nm and 675 nm-725 nm by a dual-channel view optical system (DV2, Photometrics) with a dichroic mirror (T565lpxr, Chroma) and two emission filters (ET 525/36, ET 700/50, Chroma). The split images are projected adjacently to an EMCCD camera (iXon DU-897, Andor) with a back-projected pixel size of 105 nm.

Imaging process

For each experiment, 5000 to 20000 frames of dual-channel image were typically recorded. The gain of the EMCCD was set at 150 and the exposure time was set as 50 ms for a laser power density of 5 W cm$^{-2}$ and 30 ms for a higher excitation power. Between binding experiments with different fluorescent probes on the same $MoS_2$ samples, extensive washing by the buffer solution and bleaching by a moderate laser power was performed to ensure a clean $MoS_2$ surface.

Localization procedure

Centroids of the defect-bound fluorophores are localized using the FIJI plugin ThunderStrom[52]. Briefly, we first select the image area corresponding to fluorophore emission and apply a wavelet filter (B-Spline). Fluorophores with intensity peak higher than 1.5×STD of the first wavelet level are selected for localization. The selected peaks are then fitted by 2D-Gaussian function to extract centroids. Lateral drift correction is done by cross-correlation of the reconstructed images.

Focused-ion beam patterning

FIB irradiation and patterning were performed at Helios G4 PFIB UXe microscope using focused Xenon plasma beam. All investigated patterned samples were irradiated at constant 90 μs dwell time, 2 μm pixel distance and 30 kV beam with varying beam currents (10 pA – 100 pA) for different irradiation doses. The irradiation dose was calculated with ion beam exposure formula for 2D materials[53].

Raman scattering spectroscopy

Raman spectra of the monolayer $MoS_2$ flakes were collected by a (Renishaw inVia Confocal Raman Microscope) spectrometer at room temperature using a 532 nm laser with excitation power of ~ 1 mW. Point measurements were performed on the transferred MoS$_2$ flakes on glass substrates with or without pre-patterned FIB irradiation. The excitation spot size is about 1μm.

**Data Availability**

The data that support the findings of this study are available from the corresponding authors on reasonable request.

**Acknowledgments**

We acknowledge Ilya Sychukov for valuable discussions on analytical analysis of binding kinetics. We acknowledge the support of Lely Feletti on DNA synthsis and the help of Adrien Descloux, Kristin Grussmayer and Vytautas Navikas on the fluorescence microscopy. This work was financially supported by the Swiss National Science Foundation (SNSF) Consolidator grant (BIONIC BSCGI0_157802) and CCMX project ("Large Area Growth of 2D Materials for device integration"). M.Z. acknowledges support from the Swedish Research Council through the international postdoc grant(VR 2018-06764).

**Author contributions**

A.R. supervised and coordinated all aspects of the projects. M.Z. and A.R. designed the experiments. M.Z. performed the 2D-PAINT measurements and data analysis. M.Lihter designed the chemical agent. M.M. performed FIB. K.B. performed AFM measurements. K.B. and M. Lingenfelder analyzed the AFM data. Y.Z. and J. Z did growth of MoS$_2$ and WS$_2$ under the supervision of A.K. Z.W. performed Raman spectroscopy measurements. M.Z. and J.C. wrote the manuscript with input from A.R., K.B., M.Lithter and M. Lingenfelder.